\documentclass[namedreferences,hyperref,optionalrh]{spr-sola}
\usepackage{graphicx}        
\usepackage{color}           

\usepackage{ulem}

\chardef\us=`\_

\begin{document}
\begin{frontmatter}
\title{Distribution of magnetic helicity and energy with height in solar atmosphere}

\author[addressref={aff1,aff2},email={lihongyan@bao.ac.cn}]{\inits{H.Y.}\fnm{Hongyan}~\snm{Li}\orcid{0009-0005-4168-3252}}

\author[addressref={aff1,aff2},corref,email={yangshb@nao.cas.cn}]{\inits{S.B.}\fnm{Shangbin}~\snm{Yang}\orcid{0000-0002-2967-4522}}

\author[addressref={aff1,aff2},email={xhq@bao.ac.cn}]{\inits{H.Q.}\fnm{Haiqing}~\snm{Xu}\orcid{0000-0003-4244-1077}}
\author[addressref={aff1},email={wangquan@nao.cas.cn}]{\inits{Q.}\fnm{Quan}~\snm{Wang}\orcid{0000-0003-3142-217X}}

\author[addressref={aff1},email={adxiong@nao.cas.cn}]{\inits{A.D.}\fnm{Anda}~\snm{Xiong}\orcid{0009-0002-6844-2801}}

\author[addressref={aff1,aff2},email={luopx@bao.ac.cn}]{\inits{P.X.}\fnm{PeiXin}~\snm{Luo}\orcid{0009-0000-0324-7314}}

\address[id=aff1]{State Key Laboratory of Solar Activity and Space Weather, National Astronomical Observatories, Chinese Academy of Sciences, Beijing, 100101, China}
\address[id=aff2]{University of Chinese Academy of Sciences, Beijing, 100049, China}

\runningauthor{Author-a et al.}
\runningtitle{\textit{Solar Physics} Example Article}

\begin{abstract}
Magnetic helicity and magnetic energy are key to understanding the solar dynamo and eruptions, and their three-dimensional distributions are of great significance. However, how these quantities vary with height remains poorly understood. Moreover, because the three-dimensional distribution depends on magnetic field extrapolation, determining the optimal extrapolation height from physical rather than empirical criteria remains an open problem. To address this issue, this work investigates the vertical distributions of magnetic helicity and magnetic energy in the solar corona within active regions. We analyze 150 active regions observed by the Solar Magnetic Field Telescope (SMFT) from 1988 to 2019, grouped by absolute magnetic flux, perform nonlinear force-free field (NLFFF) extrapolations, and compute the relative magnetic helicity with a finite volume method. It is found that an extrapolation height of at least 81 Mm retains 97\% of the total magnetic helicity and energy while reducing computational costs by approximately 38\% under the adopted configuration. This work provides important parameter constraints for the long-term statistical study of magnetic helicity in solar active regions.

\end{abstract}
\end{frontmatter}

\section{Introduction}
     \label{S-Introduction} 
Magnetic helicity is an important physical quantity that measures the degree of linkage and twist in the magnetic field \citep{Elsasser1956,Woltjer1958,Moffatt1969}. It is generally believed that helicity originates from the solar dynamo in the Sun’s convection zone and is transported into the atmosphere along with the emergence of magnetic flux tube \citep{Pipin1}. 
Magnetic helicity is approximately conserved in the solar atmosphere, even during fast magnetic reconnection at high magnetic Reynolds numbers
\citep{Berger1984}. 
 Magnetic helicity serves as a powerful diagnostic tool for studying the evolution of magnetic fields in the solar atmosphere \citep{HongQi1}.
 
 Common methods for calculating magnetic helicity include \citep{Valori2016}: (1) helicity-flux integration (FI); (2) discrete flux-tube (DT); and (3) finite volume (FV). 
 The FI method \citep{Chae2001} for magnetic helicity assumes that, starting from a specific initial time, the magnetic helicity accumulated within a given volume is entirely determined by the magnetic helicity flux passing through its boundary. Consequently, it requires time-series information on the temporal evolution of both the magnetic and velocity fields at the boundary of the volume under consideration (typically the bottom boundary). This implies that the method is not applicable to a single magnetogram; instead, it approximates the total coronal magnetic helicity through time integration. 
 The relative magnetic helicity can be approximated as the sum of the helicities of numerous magnetic flux tubes \citep{Berger1984,Demoulin2006}. The DT method \citep{Guo2010,Guo2013} computes magnetic helicity by summing the self helicity and mutual helicity components. However, this approach requires the prior identification of well-defined magnetic flux rope structures before it can be applied. Compared to the previous two methods, the FV method can calculate the magnetic helicity values of the magnetic field within a finite volume that satisfies certain boundary conditions. And it can calculate the data at a certain moment without considering continuous evolution. This article uses the finite volume method based on Coulomb metric developed by \citet{Yang1,Yang2}. This method requires as input the 3D magnetic field at a single moment.
 
In the nonlinear force-free field (NLFFF) model, the force-free parameter $\alpha$ is constant along each magnetic field line but may vary between different field lines. Moreover, the NLFFF model admits analytical solutions only in specific cases, such as the Low$\&$Lou mode (\citealp{Low1990}). In contrast, numerous numerical methods have been developed to solve the NLFFF problem, including the upward integration method (\citealp{Nakagawa1974,Wu1985,Wu1990}), the Grad–Rubin method (\citealp{Sakurai1981}), the MHD relaxation method (\citealp{Chodura1981,Mikic1994,Roumeliotis1996,McClymont1997}), and the optimization method (\citealp{Wheatland2000,Wiegelmann1,Wiegelmann3, Wiegelmann2008,Wiegelmann2}). The optimization method is employed for magnetic field extrapolation in this paper. 
 
Magnetic helicity is widely used to characterize the non-potentiality and eruptive potential of solar active regions. Most existing studies focus on its total value or spatial distribution at a fixed height (\citealp{Tiwari2010}). However, more recent investigations have demonstrated that magnetic helicity is not uniformly distributed on the photosphere but is highly concentrated within specific non-potential structures, such as magnetic flux ropes (\citealp{Moraitis2021}). Moreover, the evolution of topological complexity is dynamically coupled to flux emergence and submergence, exhibiting distinct spatial and temporal signatures (\citealp{MacTaggart2021}). These findings suggest that magnetic helicity is closely linked to vertically extended magnetic structures in the solar atmosphere, whereby helicity and energy are expected to show height-dependent, nonlinear accumulation and decay. 
Despite these advances, systematic studies investigating how helicity and energy accumulate with height above the photosphere remain scarce. In previous NLFFF modeling, the choice of the extrapolation ceiling in NLFFF modeling is often empirical or driven by computational constraints, lacking a physically motivated optimal cutoff height. To address the aforementioned issues, this study utilizes the long-term, stable vector magnetic field observations accumulated by the Solar Magnetic Field Telescope (SMFT) \citep{AI1986} at the Huairou Solar Observing Station (HSOS).  
This allows us to construct continuous vertical profiles of both magnetic helicity and energy from the photosphere to the corona. Specifically, we evaluate the fractional contributions of incremental height layers to the total magnetic helicity and energy by comparing nested extrapolation subvolumes, and examine their variation with height to identify a suitable extrapolation height for accurately estimating these quantities in the solar corona. 

 In section \ref{sec2}, we describe the data sources, the NLFFF extrapolation methodology, and the computational approaches for magnetic helicity and magnetic energy. In Section \ref{sec3}, we present sample magnetograms and corresponding NLFFF extrapolation results, and show the distributions of magnetic helicity and energy for active regions categorized by absolute magnetic flux. We then perform an in-depth analysis of the vertical distribution and height dependence of magnetic helicity, across groups classified by absolute magnetic flux. We also examine the vertical variations of total magnetic energy. In Section \ref{sec4}, we discuss the results from Section \ref{sec3} and propose recommended extrapolation heights for several parameters. Finally, we summarize the main findings of this study and present our conclusions.

\newpage
\section{Data and Methods}
\label{sec2}
\subsection{Data resource and preprocess}
\begin{figure}[H]
    \centering
    \includegraphics[width=1\linewidth]{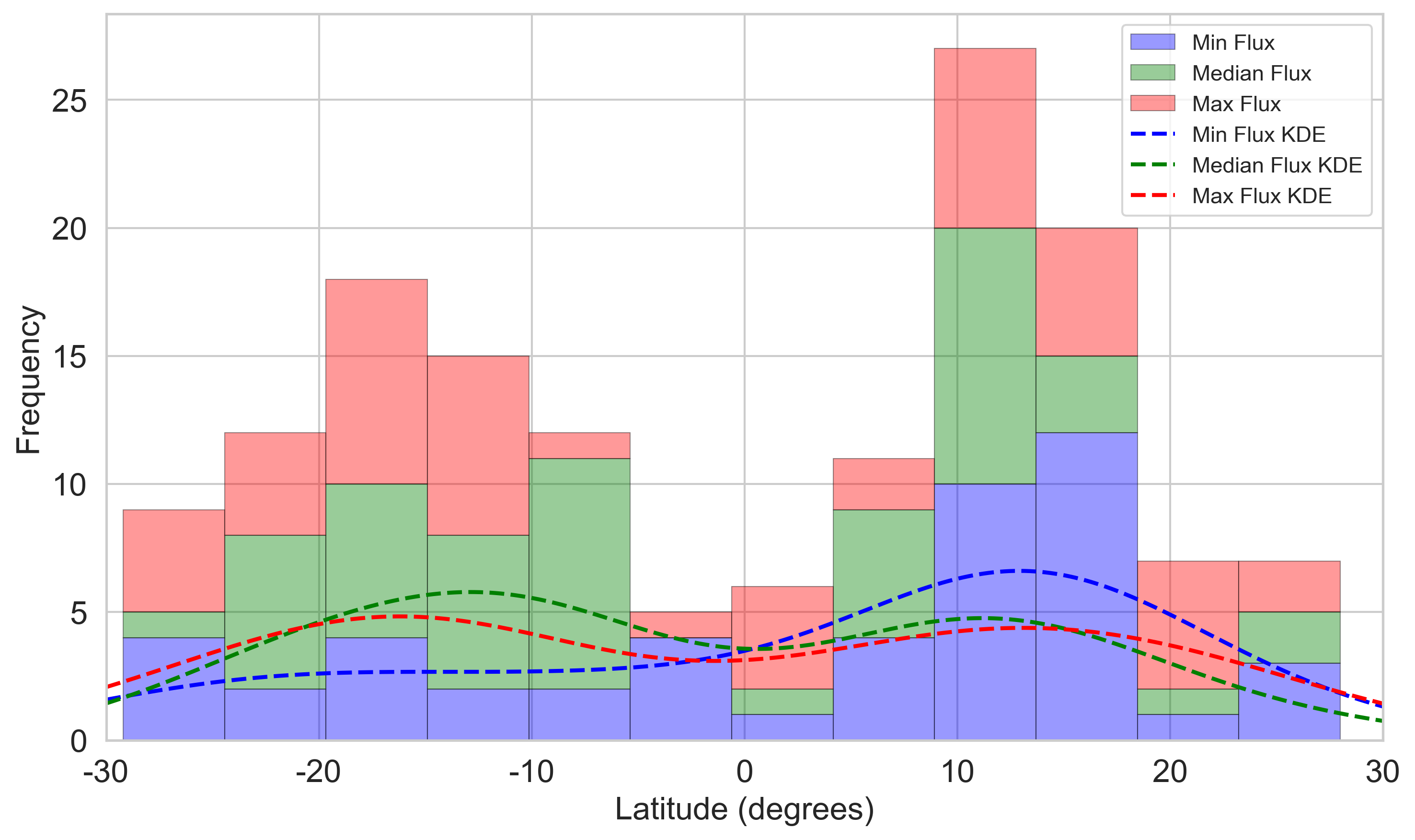}
    \caption{Where the colored bars in the figure represent the frequencies of occurrence for the three groups within the corresponding latitude bins. The dashed curves are kernel density estimation (KDE) density curves.}
    \label{1}
\end{figure} 
\noindent

Huairou Solar Observing Station, Chinese Academy of Sciences, was established in 1984. The SMFT at HSOS is designed to measure magnetic and velocity fields in solar active regions. The SMFT is equipped with a birefringent filter for wavelength selection and KD*P crystals to modulate polarization signals. It measures the magnetic field at 5324 \AA\ and the velocity field at 4861 \AA. A vector magnetogram is constructed using four narrow-band (0.125 \AA) filtergrams of the Stokes parameters $I$, $Q$, $U$ and $V$. The centre wavelength of the filter can be shifted and is normally set to $-0.075$\AA for longitudinal magnetic field measurements and to the line centre for transverse magnetic field measurements. The SMFT achieves a measurement precision of approximately 10~G for the longitudinal  magnetic field and 150~G for the transverse component, with a temporal resolution of approximately 3--5 minutes for vector magnetograms \citep{AI1986,Su2024}. Through these long-term SMFT observations, HSOS has collected a valuable dataset of vector magnetic fields covering nearly four solar cycles.

We calculated absolute magnetic flux ($\Phi_{abs}$) for a dataset of 58,942 SMFT magnetograms spanning 1988–2019, covering 3,248 solar active regions. From this dataset, we select magnetograms according to three selection criteria: 
(1) We only select data with heliocentric angle $\theta \le$ 30 degrees to minimize projection effects;
(2) For multiple observations of the same active region, we retain the magnetogram with the maximum $\Phi_{abs}$ and the lowest noise level;
Subsequently, we selected 150 magnetograms (each magnetogram has only one active region) and categorize them into three groups: 50 with the highest flux (max flux; $\Phi_{abs} = 4.12-8.14\times 10^{22}$ Mx), 50 with median flux (median flux; $\Phi_{abs} = 1.33-1.41\times 10^{22}$ Mx), and 50 with the lowest flux (min flux; $\Phi_{abs} = 2.80-6.15\times 10^{21}$ Mx).The distribution of flux for all 150 magnetograms is shown in Figure \ref{1}, where colored blocks represent different flux groups and the kernel density estimation (KDE) curve shows the flux distribution. Before extrapolation and magnetic helicity calculation, we perform the following preprocessing steps: (1) polarization crosstalk correction (\citealp{sujiangtao}); (2) removal of the Faraday rotation effect (\citealp{Gaoyu1,Gaoyu2}); and (3) 180° ambiguity resolution (\citealp{WangT}).

\subsection{Extrapolation method}
Reliable, high-accuracy magnetic field measurements are only available in the photosphere. However, we can use extrapolation methods to infer the magnetic field configuration in the upper solar atmosphere. In this paper, we apply the NLFFF model \citep{wiegelmann2012,Wiegelmann2} to extrapolate the 3D magnetic field from observed photospheric vector magnetograms. This model is based on two fundamental assumptions: the magnetic field is divergence-free ($\nabla\cdot \mathbf{B}=0$), and in the solar corona, the field is force-free ($\mathbf{J}\times\mathbf{B}=0$) \citep{Wiegelmann2021},  This method employs an optimization algorithm to minimize an objective function:

\begin{equation}
L=\int_V \omega(x,y,z)[\mathbf{B}^{-2}|(\nabla \times \mathbf{B})|^{2}+|\nabla \cdot \mathbf{B}|^{2}]d^3x
\end{equation}
\noindent
where $\omega(x,y,z)$ is a weight function used to minimize the influence of the boundary. It is set to 1 in the core region and gradually decreases to zero near the boundary. The boundary layer thickness is typically set to 10--20 grid points to balance the stability of the solution and the suppression of boundary effects. Since real observational data do not satisfy the force-free condition, we follow the method of \citet{Wiegelmann3,Schrijver} to preprocess all selected vector magnetic field data. Within the NLFFF method, we employ a multigrid method \citep{MacTaggart,Gilchrist,Zhu} for numerical implementation. At the same resolution, the computation time is approximately half that of direct calculation. Moreover, the method enhances the stability of the extrapolation results by progressively refining the grid from coarse to fine during optimization, thereby facilitating global convergence.

Once the 3D magnetic field is extrapolated from a vector magnetogram, it is essential to evaluate how well the solution satisfies the force-free and divergence-free constraints of the NLFFF model. According to \citet{Low1985}, the magnetic field is force-free when:
\begin{equation}
|F_x|\ll |F_p|,\quad  |F_y|\ll |F_p|,\quad |F_z|\ll |F_p|
\end{equation}
where $F_x$, $F_y$, and $F_z$ are the components of the net Lorentz force, and $F_p$ is the characteristic magnitude of the total Lorentz force. \citet{Metcalf1995} proposed that when $|F_x/F_p|$, $|F_y/F_p|$, and $|F_z/F_p|$ are all less than 0.1, the magnetic field can be considered a force-free field. Moreover, three widely used parameters are employed to quantify the quality of the extrapolation and magnetic helicity results: $\sigma_J$, $\langle|f_i|\rangle$, $E_{\rm div}/E$. Here $\sigma_J$ is the weighted average of the angle between the magnetic field and the current, used to quantify the force-free nature of the magnetic field (\citealp{Wheatland2000}); $\langle|f_i|\rangle$ is the fractional magnetic flux, used to characterize the divergence-free property of the field (\citealp{Valori,Thalmann}); And $E_{\rm div}/E$ provides a metric for assessing the reliability of the magnetic helicity calculation (\citealp{Thalmann,Moraitis2024}).

\subsection{Magnetic helicity calculation method}
The magnetic helicity is defined by:

\begin{equation}
H_M=\int_V\mathbf{A}\cdot \mathbf{B}dV
\label{m4}
\end{equation}
\noindent
where $\mathbf{A}$ is the magnetic vector potential and $\mathbf{B}$ is the magnetic field. This equation measures the twist and writhe of the magnetic field in a finite volume. However Equation~\ref{m4} has a fundamental limitation: $\mathbf{A}$ cannot be directly observed or measured, and it lacks gauge invariance, which is crucial for the accurate calculation of magnetic helicity. Therefore, $H_M$ is well-defined value only when the normal component of the magnetic field vanishes on the boundary of the closed volume, i.e., $\mathbf{B}\cdot \hat{\mathbf{n}}|_S =0$. To address this issue, \citet{Berger1984} introduced the concept of relative magnetic helicity, which is defined as:

\begin{equation}
H_R=\int_V(\mathbf{A}+\mathbf{A}_P)\cdot (\mathbf{B}-\mathbf{P})dV
\label{eq4}
\end{equation}
\noindent
 where $\mathbf{A}_P$ is the vector potential of the potential field, also known as the current-free field, which satisfies $\nabla\times \mathbf{P}=0$. Compared to $H_M$ in Equation \ref{m4}, $H_R$ quantifies the magnetic helicity of the actual field $\mathbf{B}$ relative to the reference potential field $\mathbf{P}$. This relative formulation ensures gauge invariance and allows application to a finite volume with arbitrary boundary conditions.

 In this paper, we apply Coulomb-Yang method \citep{Yang1,Yang2} to compute the relative magnetic helicity from the extrapolated magnetic field.  The method first takes the three-dimensional magnetic field distribution $\mathbf{B}(x,y,z)$ within a finite volume as input, and then constructs a potential field $\mathbf{P}$ that matches the normal component of $\mathbf{B}$ on the boundary, i.e.,

\begin{equation}
 \mathbf{P}\cdot \hat{\mathbf{n}}|_S=\mathbf{B}\cdot \hat{\mathbf{n}}|_S
\end{equation}
\noindent
thereby making:
 
\begin{equation}
 (\mathbf{B-P})\cdot \hat{\mathbf{n}}|_S=0
\end{equation}
\noindent
then utilize definition formula of vector potential to calculate $\mathbf{A}_P,\mathbf{A}$ and employing the Coulomb gauge:

\begin{equation}
\nabla\times\mathbf{A}=\mathbf{B} , \nabla\times\mathbf{A}_P=\mathbf{P}
\end{equation}

\begin{equation}
\nabla\cdot\mathbf{A}=0,\nabla\cdot\mathbf{A}_P=0
\end{equation}
\noindent
at this moment, we can substitute $
\mathbf{A}, \mathbf{A}_P,\mathbf{P},\mathbf{B}$ into Equation \ref{eq4} and calculate relative magnetic helicity.

To study about magnetic helicity and energy evolution with height, we sliced the extrapolation results at intervals of every 5 layers in height, and we perform only one 3D NLFFF extrapolation for each magnetogram. To obtain the magnetic helicity at different heights, we do not re-extrapolate the field; instead, we slice the extrapolated 3D field to define specific subvolumes.
\begin{figure}[H]
    \centering
\includegraphics[width=0.6\linewidth]{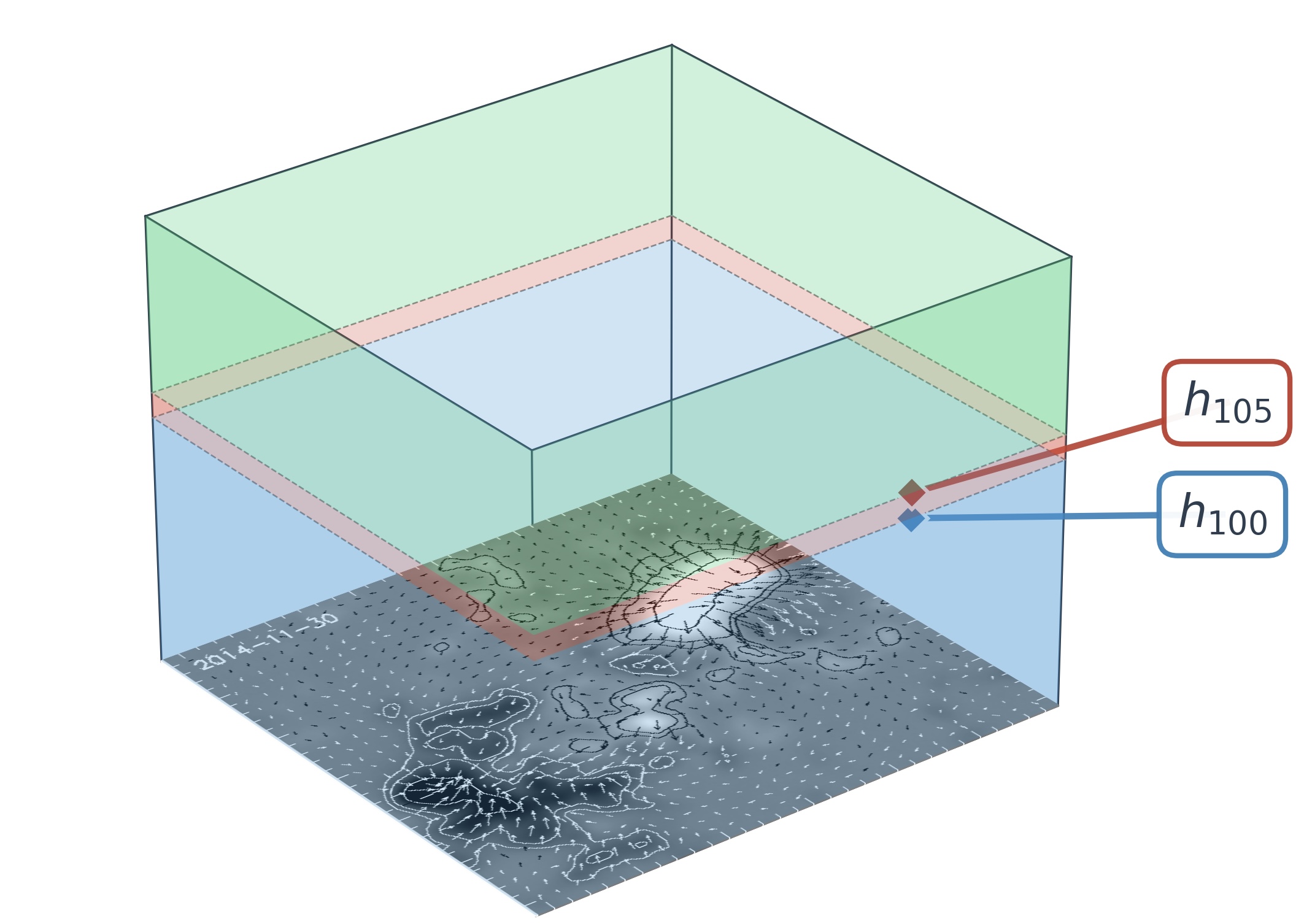}
    \caption{Schematic diagram of extrapolated slices used to compute magnetic helicity at different heights.}
    \label{2}
\end{figure} 
\noindent

 As shown in Figure \ref{2}, to quantify the vertical distribution of magnetic helicity, we evaluate the fractional contribution of specific height layers by comparing nested subvolumes. Taking $0 \le h \le 100$ and $0 \le h \le 105$ as representative cases, we extract the magnetic field data within each target subvolume from the extrapolated 3D field. For the range $0 \le h \le 100$ (blue region), we compute the corresponding potential field and the vector potential to calculate the magnetic helicity as shown in Equation \ref{eq4}. The same procedure is applied to the extended range $0 \le h \le 105$ (combined blue and red regions). The contribution of an incremental layer $[100, 105]$ is then estimated as $\Delta H = H(0 \le h \le 105) - H(0 \le h \le 100)$. Normalizing this difference by the helicity of the reference volume ($H_{\rm total}$; i.e., the magnetic helicity derived from the entire extrapolated 3D magnetic field distribution, without subvolume partitioning) yields the fractional contribution $f = \Delta H / H_{\rm total}$. We emphasize that, due to the non-additivity of relative magnetic helicity \citep{Valori2020}, $\Delta H$ does not  equal the helicity contained in the isolated layer $[100, 105]$. This approach therefore serves as a numerical estimate, adopted purely to facilitate percentage-based comparison and discussion of the height-dependent helicity distribution.

\subsection{Calculation method for Magnetic energy}
Three-dimension vector magnetic field extrapolate result include $B_x$, $B_y$, $B_z$. It is simple to calculate magnetic energy by use this formula \citep{Vemareddy2024}:

\begin{equation}
\mathbf{E}_{total}=\frac{1}{8\pi}\int_V \mathbf{B}^2dV
\end{equation}
\noindent
where $\mathbf{B}^2=\mathbf{B}_x^2+\mathbf{B}_y^2+\mathbf{B}_z^2$.
\newpage

\section{Results}\label{sec3}
\begin{figure}[H]
    \centering
    \begin{minipage}{0.482\textwidth}
        \centering
        \includegraphics[width=\linewidth]{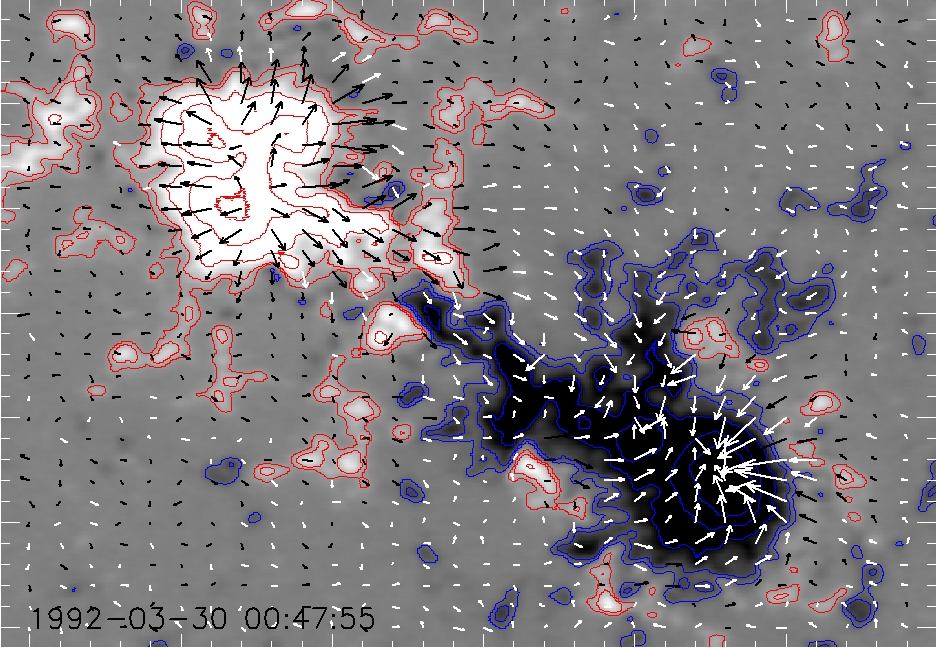}
    \end{minipage}
    \hfill
    \begin{minipage}{0.50\textwidth}
        \centering
        \includegraphics[width=\linewidth]{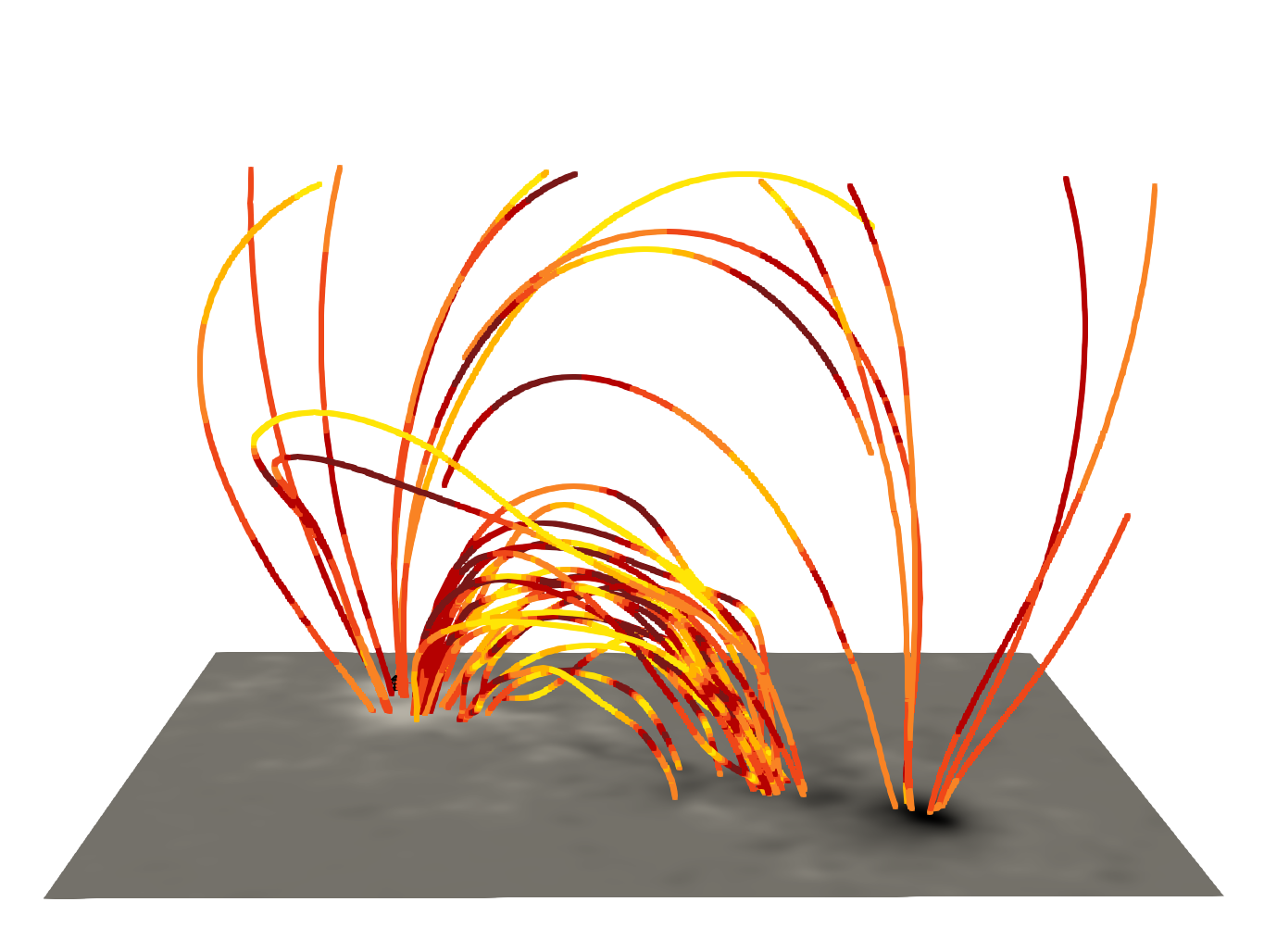}
    \end{minipage}
    \caption{Where the left-hand is magnetograph observed at 1992/03/30 and belong to max group. On the right-hand side is a schematic illustration of the corresponding NLFFF extrapolation result.}
    \label{4}
\end{figure}

Examples of the extrapolation results and their corresponding magnetograms are shown in Figure \ref{4}. The left panel displays the magnetogram from the max-flux group, while the right panel shows the corresponding extrapolated 3D magnetic field. Before extrapolation, we resampled all magnetograms to a uniform resolution of 1~arcsecond to ensure that the magnetic helicity and magnetic energy were computed on the same spatial scale. We use a multigrid order of 3. The initial extrapolation was performed at a coarsened resolution of 4~arcsecond (i.e., one-quarter of the original magnetogram resolution), which simplifies the computation and improves the stability of the extrapolation results \citep{Wang}.
We set the boundary layer of the extrapolation box to 16 grid points and the extrapolation height to 200 grid points, resulting in a physical extent of approximately 133~Mm, which encompasses the majority of the corona and its key magnetic structures.  We also computed a set of quality parameters for all 150 extrapolation results. The $|F_x/F_p|$, $|F_y/F_p|$ and $|F_z/F_p|$ are all less than 0.1, indicating a well-constructed force-free magnetic field. The median values of $\langle |f_i| \rangle\times 10^4$ and $\theta_J$ ($\theta_J=\arcsin(\sigma_J)$) are 4.09 (IQR: 2.83--5.75) and 3.59 (IQR: 2.87--5.31) degrees, respectively, while $E_{\rm div}/E$ has a median value of 0.0159 (IQR: 0.0092--0.0333); here IQR represents the interquartile range. This value are lower than the 0.05 suggested by \citet{Thalmann}, indicating meaningful and reliable magnetic helicity computations.

\begin{figure}[H]
    \centering
    \begin{minipage}{0.495\textwidth}
        \centering
        \includegraphics[width=\linewidth]{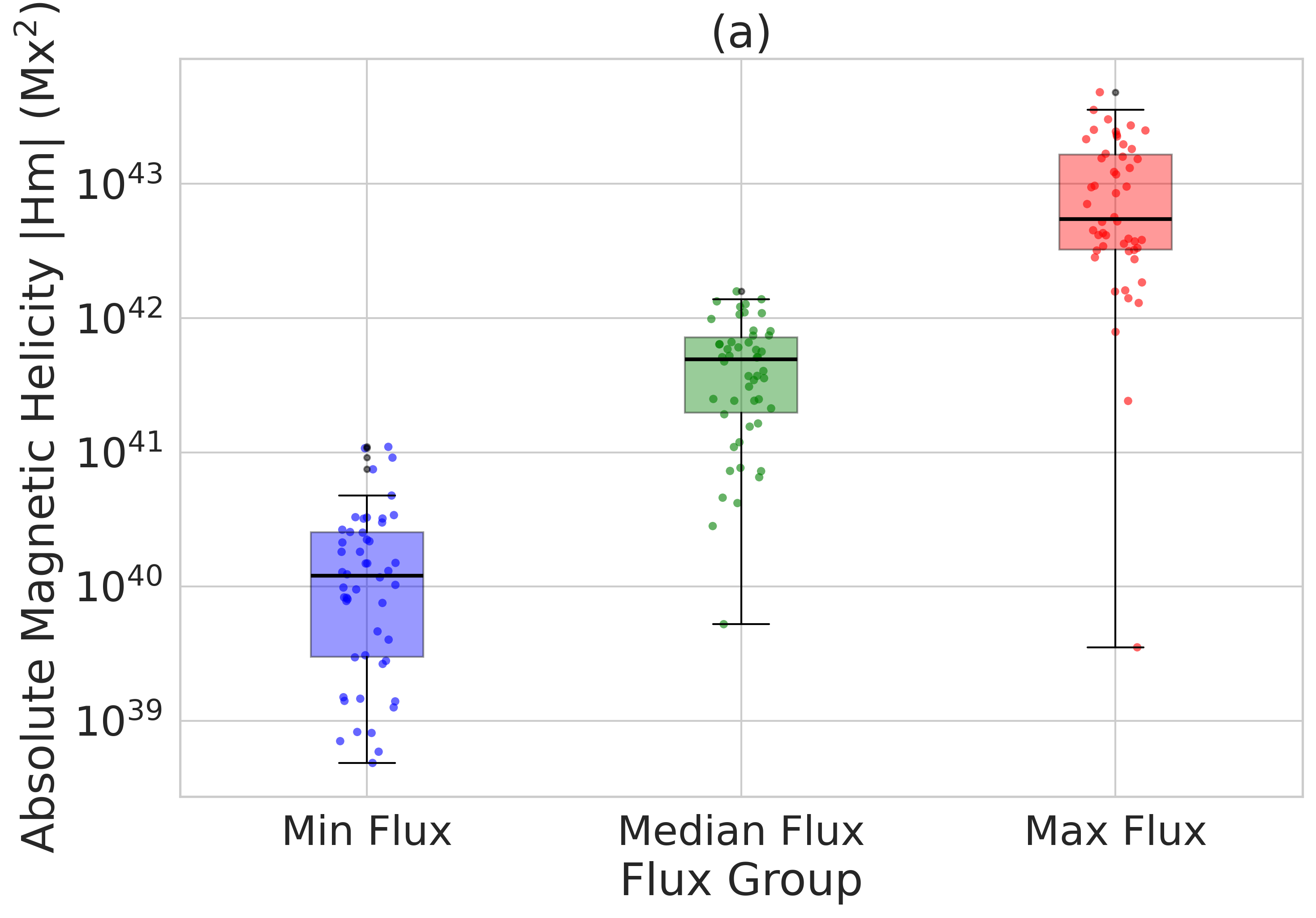}
    \end{minipage}
    \hfill
    \begin{minipage}{0.495\textwidth}
        \centering
        \includegraphics[width=\linewidth]{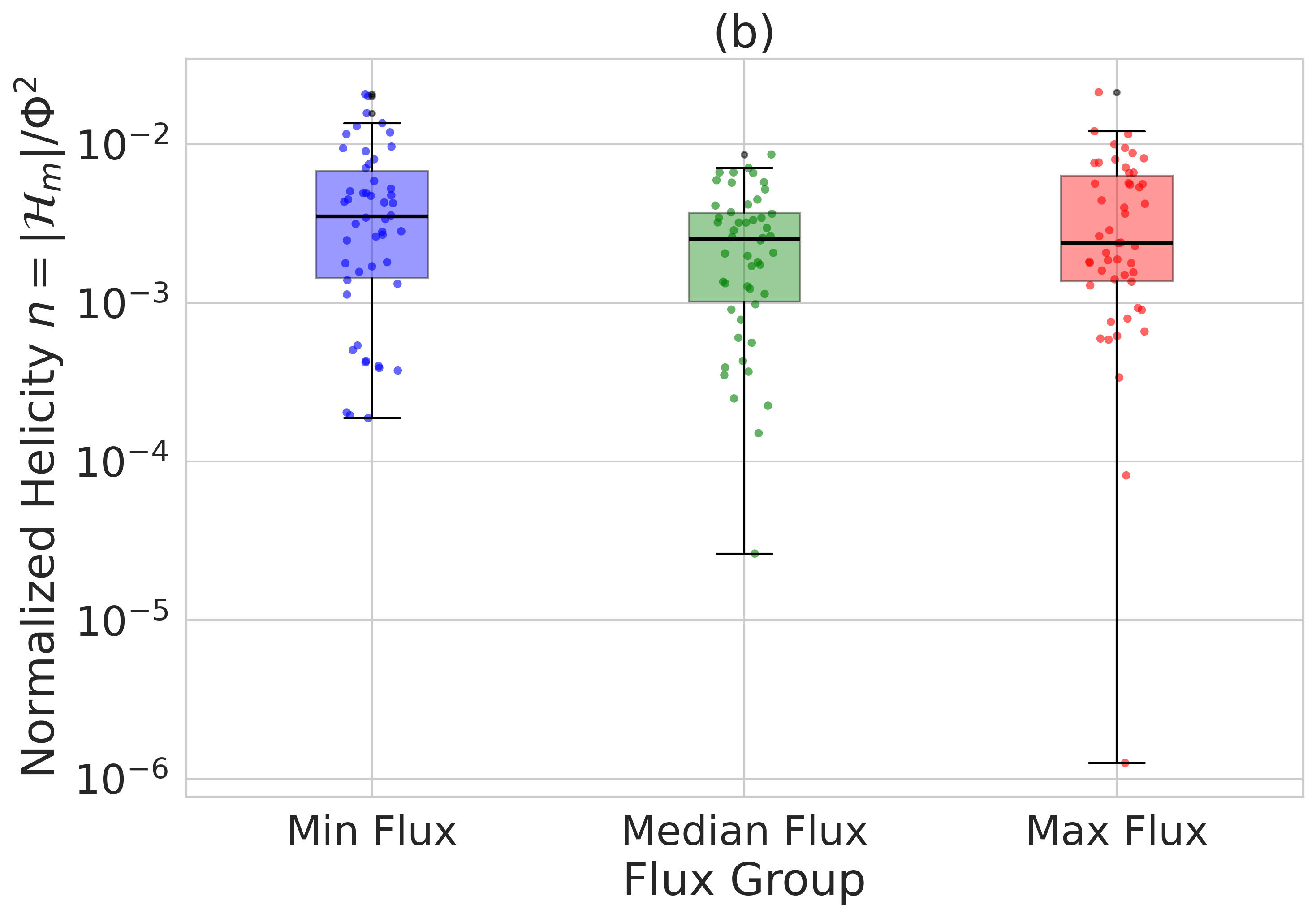}
    \end{minipage}
    \vspace{1em} 
    \begin{minipage}{0.495\textwidth}
        \centering
        \includegraphics[width=\linewidth]{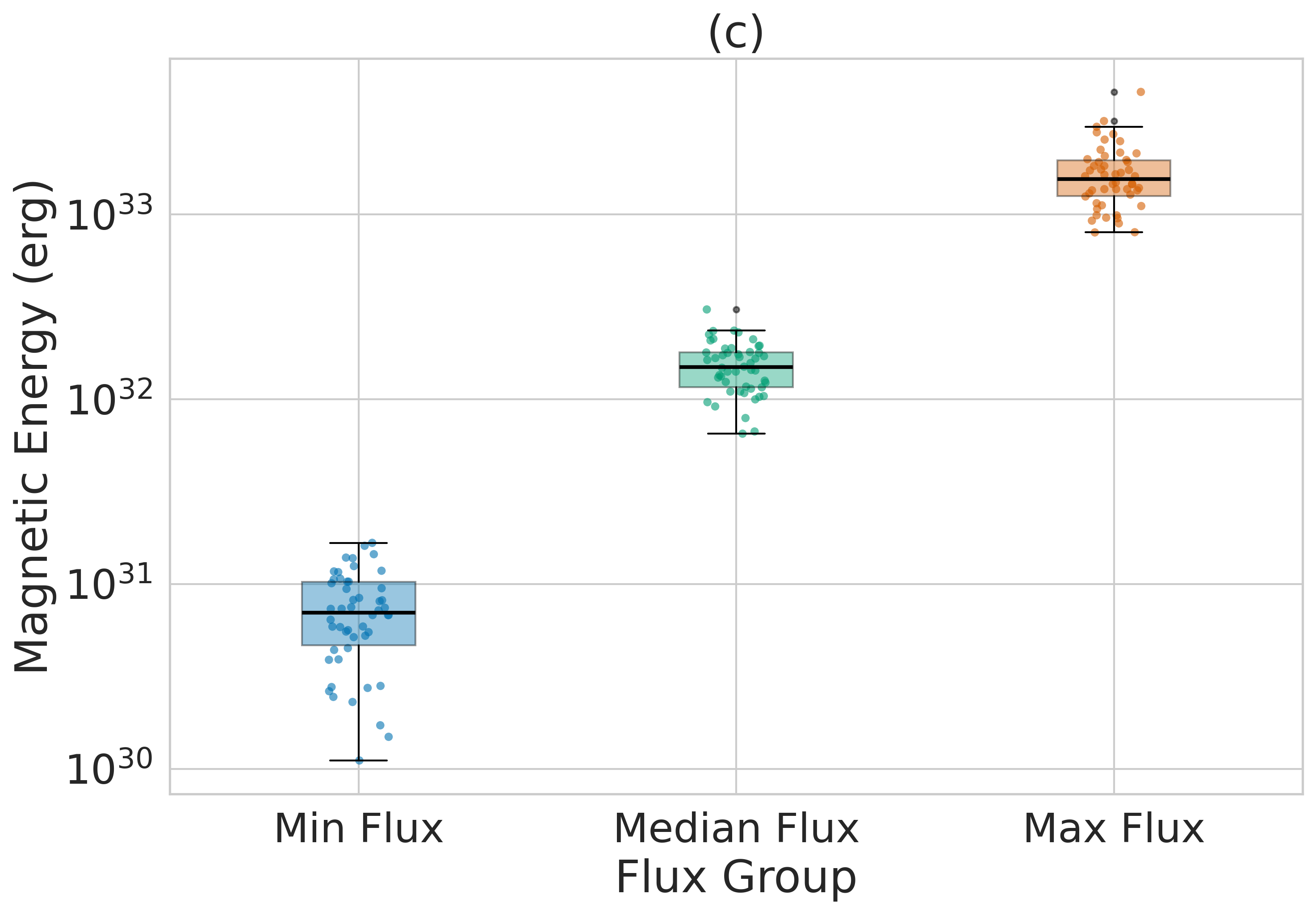}
    \end{minipage}
    \hfill
    \begin{minipage}{0.495\textwidth}
        \centering
        \includegraphics[width=\linewidth]{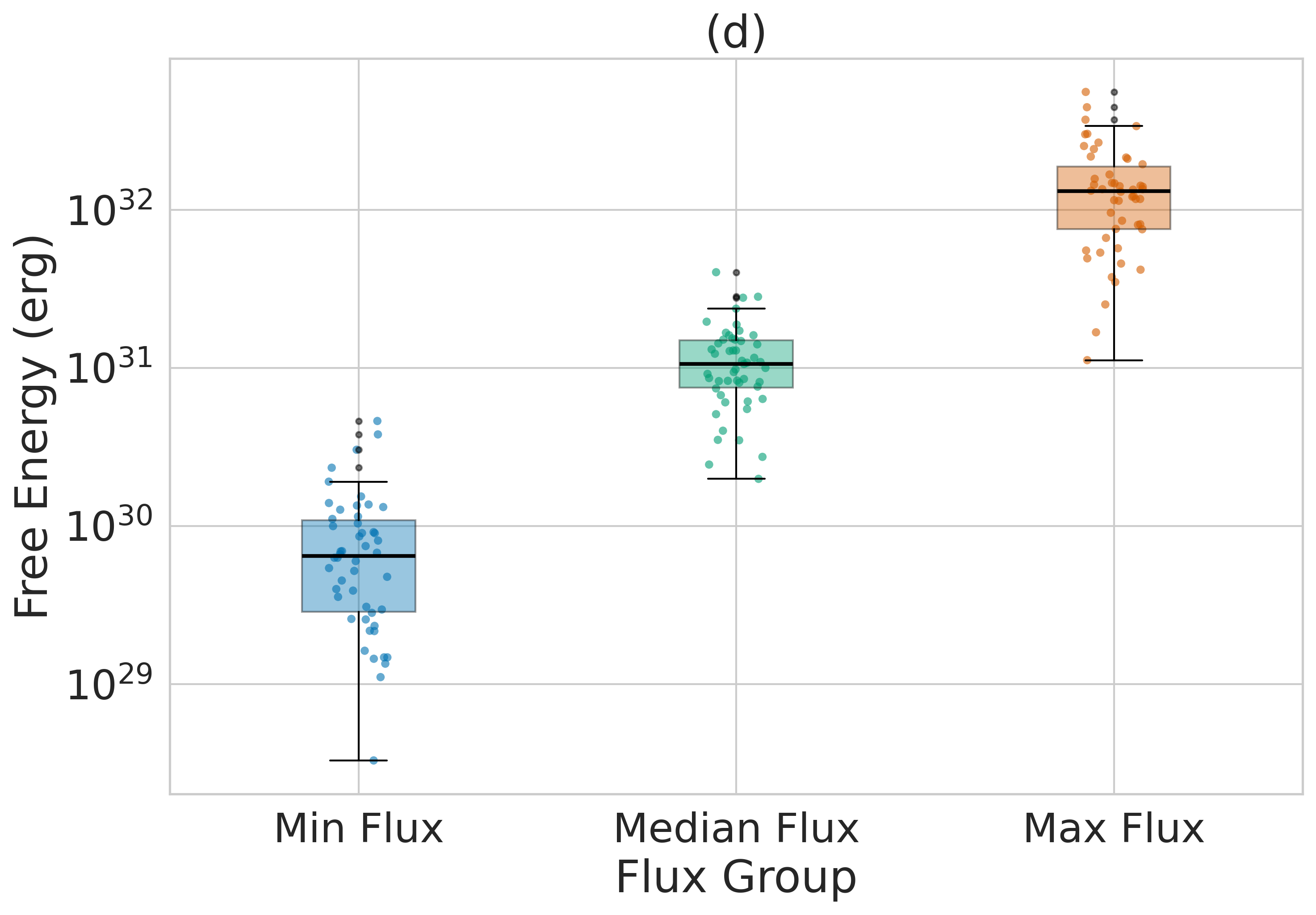} 
    \end{minipage}
    
    \caption{Panel (a) shows the distribution of absolute magnetic helicity, panel (b) the normalized magnetic helicity distribution, panel (c) the total magnetic energy distribution, and panel (d) the free energy distribution. Where the semi-transparent colored blocks in the plot represent the interquartile range (IQR), i.e., the middle 50\% of the data, and the thick black horizontal line inside each block indicates the median.}
    \label{5}
\end{figure}

In Figure \ref{5}, we display the distributions of the different groups of absolute helicity, helicity normalized by $\Phi_{\rm abs}^2$, magnetic energy and free energy. The two panels in the first row show the absolute magnetic helicity and the normalized helicity for the three different flux groups. The second row shows the magnetic energy and free energy for the three different flux groups. The max group exhibits helicity magnitudes of $\sim 10^{43} \mathrm{Mx}^{2}$, while the median and min groups reach $\sim 10^{41} \mathrm{Mx}^{2}$ and $\sim 10^{40} \mathrm{M}x^{2}$, respectively, reflecting their differing magnetic flux levels. We find that the absolute helicity, magnetic energy, and free energy increase with magnetic flux, whereas the normalized helicity shows no significant variation across groups and exhibits good consistency. This result is consistent with previous studies \citep{LaBonte2007,Yang2009} and indicates that magnetic flux tubes in large or small active regions are similarly twisted. 

For each group,we computed the variation of absolute magnetic helicity with extrapolation height for every magnetogram and then averaged the results across the group. We also evaluated the fractional contribution of magnetic helicity at each height layer to the total volume. 

\begin{figure}[H]
    \centering
    \includegraphics[width=1\linewidth]{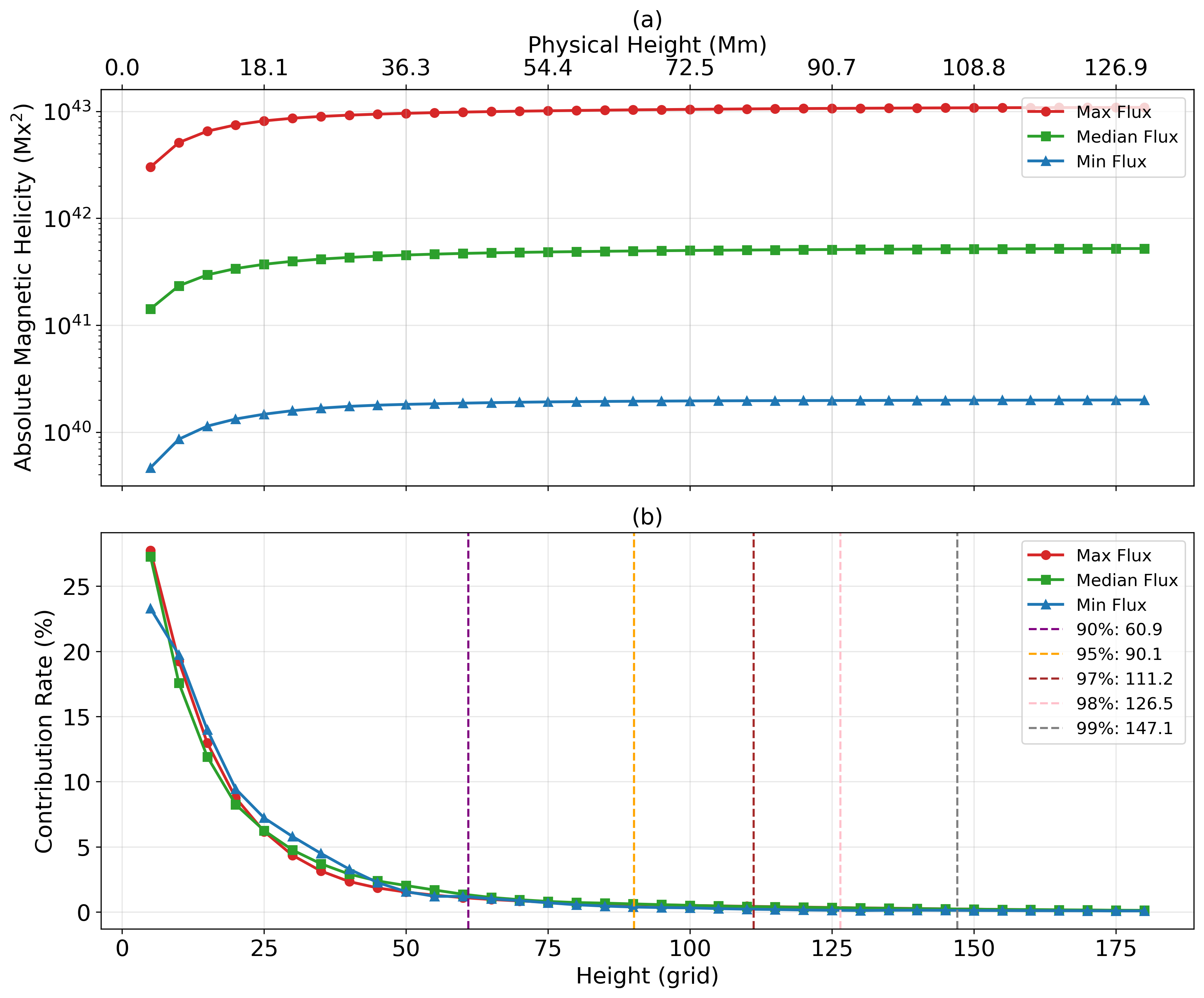}
    \caption{Panel (a) shows the absolute magnetic helicity, and panel (b) shows its contribution (i.e., the percentage of total helicity) as a function of height. Where the red, green, and blue curves represent the max, median, and min groups, respectively. The dashed vertical lines in different colors indicate the height where the corresponding contribution rate is achieved.}
    \label{6}
\end{figure}

In Figure \ref{6}, the top row plots the absolute magnetic helicity as a function of height for each group, while the bottom row shows the fractional contribution $f$ of each height layer to the total helicity. The magnetic helicity is predominantly concentrated at lower heights. Specifically, height levels up to 60.9, 90.1, 111.2, 126.5, and 147.1 grid units contribute 90\%, 95\%, 97\%, 98\%, and 99\% of the total helicity, respectively. Notably, the first 61 height layers account for 90\% of the total magnetic helicity, while the remaining 119 layers (from height 61 to 180) contribute only 10\%. This strongly suggests that the majority of magnetic helicity is stored in the lower corona, with diminishing contribution from higher heights. It is therefore evident that, in the magnetic field extrapolation, the marginal gain in magnetic helicity computation diminishes with increasing height. This exhibits a clear diminishing-returns effect. Considering that the discrepancies among FV methods for computing magnetic helicity are typically within 3\% \citep{Valori2016}, we propose that the height  $h_{\rm helicity}$ , at which the cumulative helicity reaches 97\% of its total, serves as a practical and sufficient upper boundary for helicity calculations within acceptable error tolerances. This height corresponds to a grid height 111 (physical height 81 Mm).  Extrapolating beyond this height yields negligible improvement in helicity accuracy while significantly increasing computational cost, and thus can be reasonably omitted in numerical implementations.

\noindent

\begin{figure}
    \centering
    \begin{minipage}{0.495\textwidth}
        \centering
        \includegraphics[width=\linewidth]{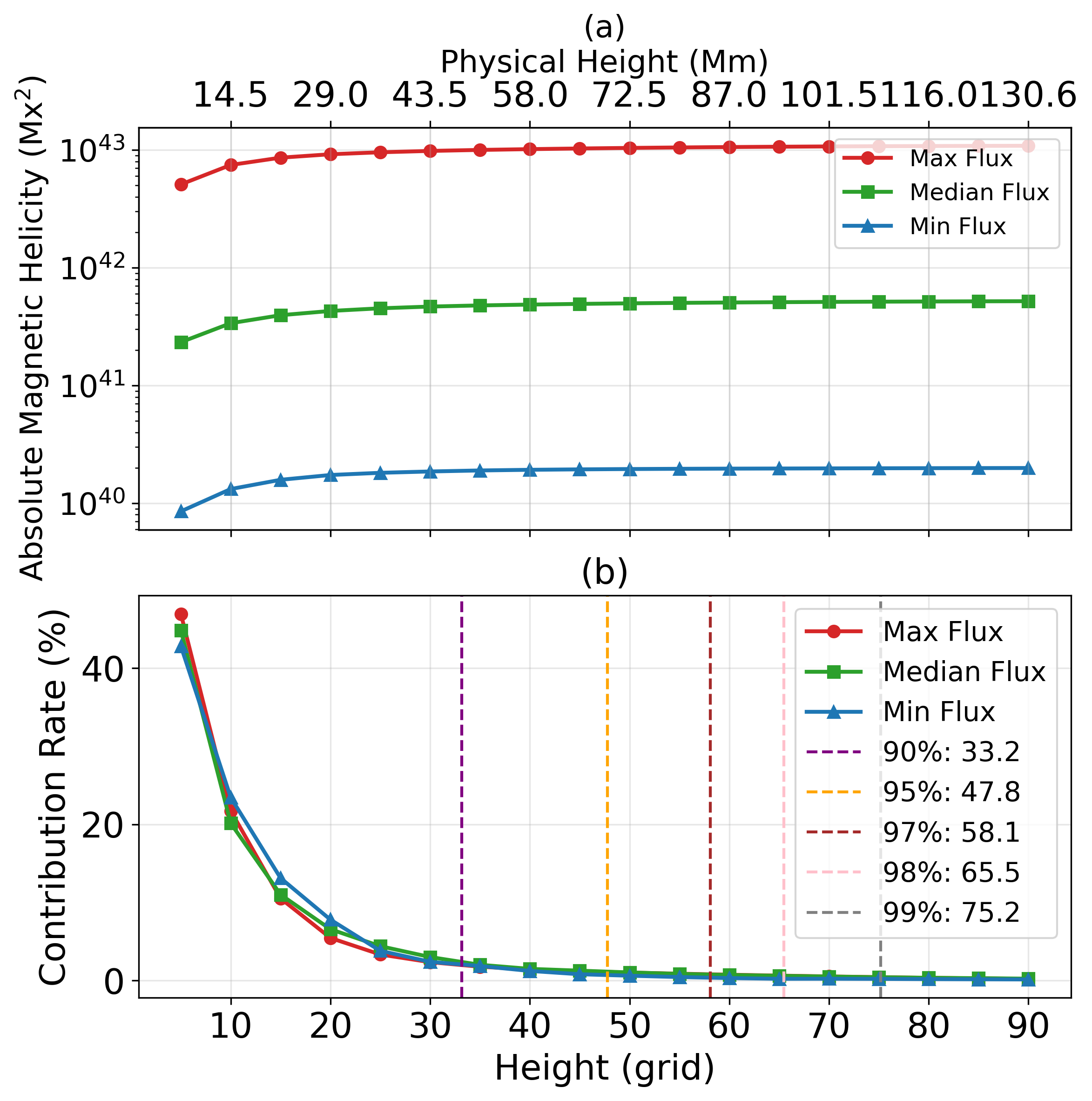}
    \end{minipage}
    \hfill
    \begin{minipage}{0.495\textwidth}
        \centering
        \includegraphics[width=\linewidth]{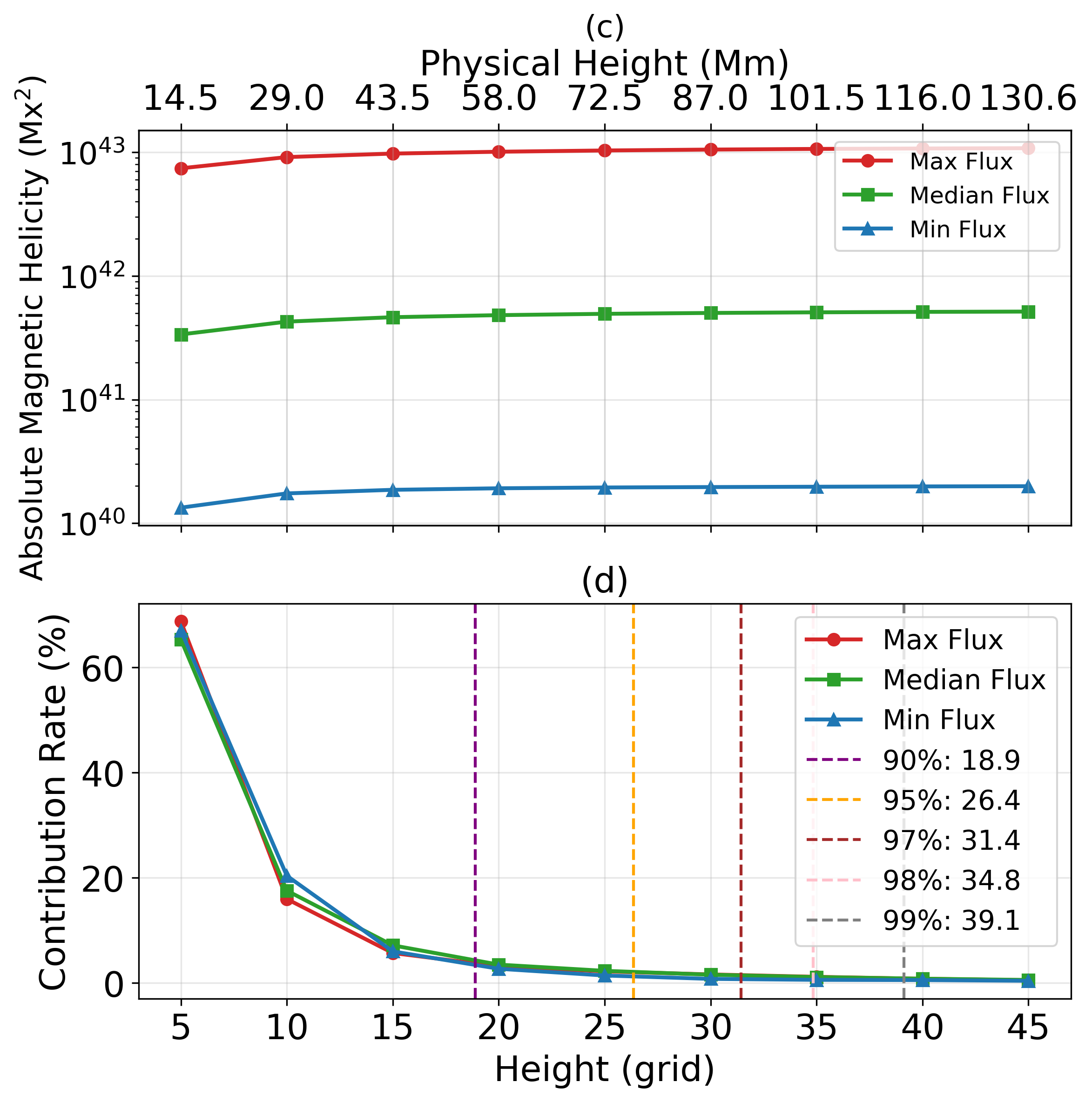}
    \end{minipage}
    \caption{Panels (a) and (b) show the absolute magnetic helicity and its fractional contribution (i.e., the percentage of total helicity) as a function of height for the magnetogram at 1/2 original size, respectively. Panels (c) and (d) present the same quantities for the magnetogram at 1/4 original size. In all panels, the red, green, and blue curves represent the max, median, and min groups, respectively. The dashed vertical lines in corresponding colors mark the heights at which the indicated contribution rates are achieved. }
    \label{7}
\end{figure}

\begin{figure}[H]
    \centering
    \includegraphics[width=1\linewidth]{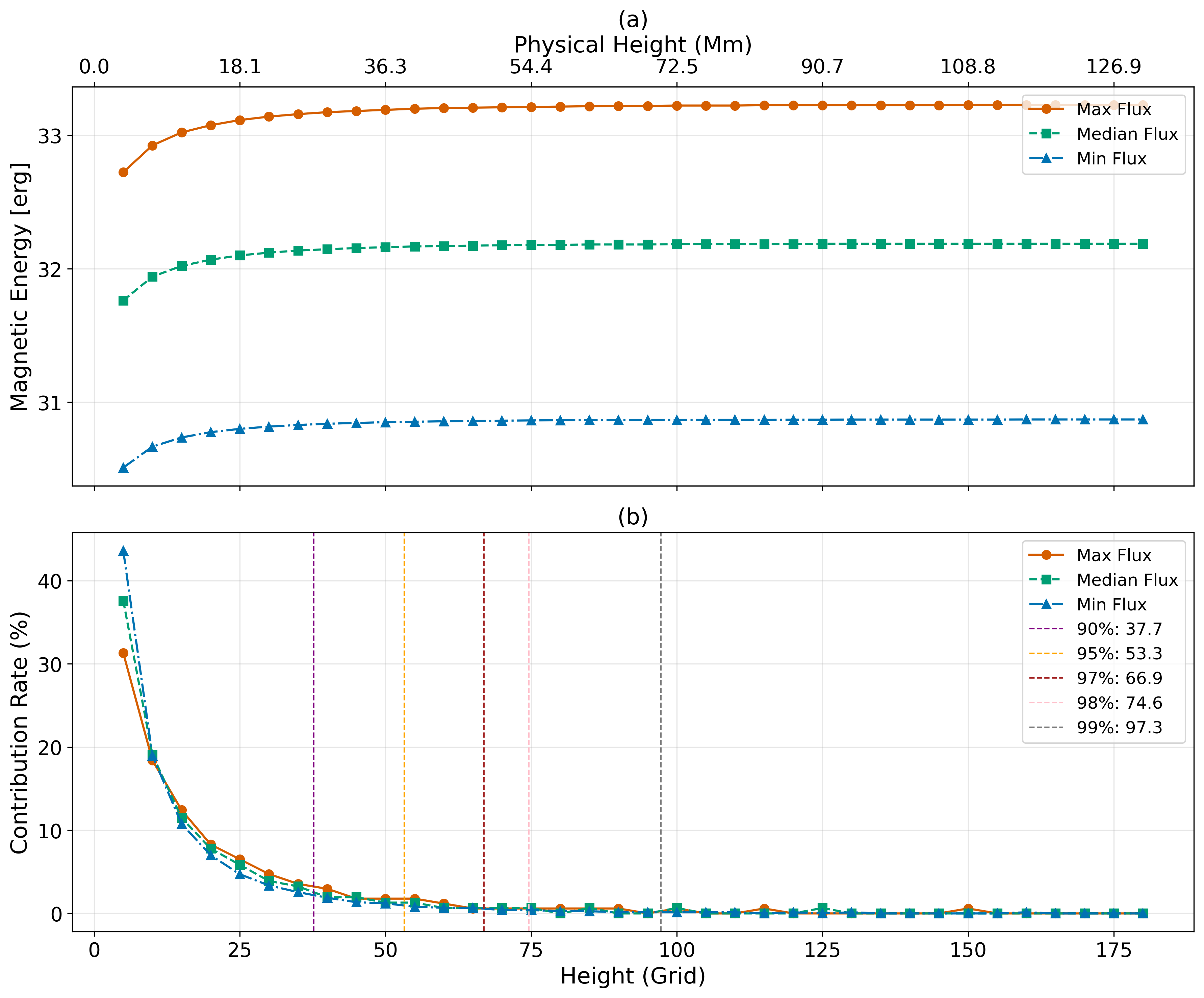}
    \caption{Panel (a) shows the height-averaged cumulative total magnetic energy, and panel (b) shows its fractional contribution (i.e., the percentage of total energy) as a function of height. Where the light orange, light green, and light blue lines represent the magnetic energy of the max, median, and min groups, respectively. The dashed vertical lines in different colors indicate the height where the corresponding contribution rate is achieved.}
    \label{8}
\end{figure}

In addition, to investigate the influence of magnetogram size on the results, we repeated the above magnetic helicity calculations and analyses using data scaled to 1/2 (pixel resolution is 2 arcseconds) and 1/4 (pixel resolution is 4 arcseconds) of the original dimensions, respectively. The resulting comparisons are presented in Figure \ref{7}. As shown, the magnetic helicity computed at reduced resolutions, similar to that shown in Figure \ref{6}, is predominantly concentrated at lower heights. Moreover, the grid heights at which the cumulative magnetic helicity reaches 97\% of the total helicity are 58.1 and 31.4 for the one-fold and two-fold downsized cases, respectively, corresponding to physical heights of approximately 84.3 Mm and 91 Mm, which are slightly higher than the recommended height of $h_{helicity} = 81$ Mm mentioned earlier. This discrepancy may arise because reducing the resolution suppresses noise to some extent, thereby leading to subtle differences in the calculated results.     
In addition to magnetic helicity, we also examine the vertical distribution of magnetic energy and free energy, and propose a reference height based on their behavior with height. So we also calculate contribution rate with height of magnetic energy and free energy as show in Figure \ref{8}. 

It can be seen that magnetic energy is also predominantly concentrated at lower heights. Specifically, grid heights 38, 53, 67, 75, and 97 account for 90\%, 95\%, 97\%, 98\%, and 99\% of the total magnetic energy, respectively. If we adopt the same 97\% contribution threshold used for helicity calculations, the recommended extrapolation heights for magnetic energy is grid heights 67, corresponding to physical heights of 49 Mm, respectively. When the extrapolation reaches or exceeds these heights, the error in the computed energy will be less than 3\%.

\section{Discussion and Conclusions}\label{sec4}
Based on the SMFT dataset, we grouped the magnetograms according to their absolute magnetic flux and selected 150 representative cases. (Max, Median, and Min groups).  Subsequently, we performed NLFFF extrapolations to compute magnetic helicity and magnetic energy. We then investigated the vertical distribution and evolution of these quantities with height in the solar atmosphere. In Section \ref{sec3}, we proposed recommended extrapolation heights for the reliable computation of these parameters. Specifically, we recommend 81~Mm for magnetic helicity and 49~Mm for magnetic energy as reference lower bounds. Extrapolating to these heights ensures approximately 97\% retention of the total magnetic helicity and energy within acceptable error tolerances. Modestly exceeding these values is permissible; however, further extrapolation yields negligible improvements in accuracy while significantly increasing computational costs. Conversely, extrapolating below these thresholds results in insufficient accuracy. This balance between reliability and efficiency reduces computational costs by approximately 38\% compared to the full extrapolation height of 133~Mm used in this study.

It is important to note that the typical physical dimensions of the magnetograms used in our study are approximately 190 $\times$ 130~Mm. At this scale, the field of view is sufficient to encompass the vast majority of solar active regions, with the exception of only a few exceptionally large ones. Based on these physical dimensions, we derived recommended extrapolation heights of 81 Mm for magnetic helicity calculations and 49 Mm for magnetic energy. If larger-dimensional data were used, which would include more quiet-Sun regions beyond the active region, it can be anticipated that the results would not be significantly affected. This is because the magnetic fields in quiet-Sun regions are relatively weak, making their contributions to both magnetic energy and helicity negligible. On the other hand, as the resolution decreases, the height corresponding to 97\% contribution increases slightly. For instance, when the resolution is reduced by a factor of 2, this height increases to 84.3 Mm, and when reduced by a factor of 4, it reaches 91 Mm. This implies that for lower-resolution magnetograms, the extrapolation height could be appropriately increased to maintain comparable accuracy.

Filaments, or prominences when observed at the solar limb, are structures composed of relatively cool and dense plasma in the solar atmosphere, typically suspended in the corona. The number of filaments decreases exponentially with height in the range of 10–100 Mm, with an average height of 30 Mm \citep{Ananthakrishnan1961,Filippov2013}. More importantly, \citet{Filippov2013} demonstrated that a quiescent filament erupted precisely when its height reached the critical value of 80 Mm. Similarly, \citet{Cheng2020} found that the onset of the main acceleration phase in quiescent filament eruptions typically occurs at an average height of 118 Mm, significantly higher than that of active-region hot channels (near 50 Mm). Moreover, recent NLFFF modeling of a 500 Mm-long filament rooted in weak-field regions (\citealp{Purkhart2025}) revealed that the erupting magnetic flux rope reaches heights of 60~Mm, while the overlying strapping field extends beyond 100 Mm. Therefore, if the extrapolation is performed only up to relatively low heights, critical information related to the key structures of filaments and filament eruptions may be missed. Hence, 81~Mm represents a practical and sufficient cut‑off height for helicity calculation within acceptable error tolerances, which also aligns well with the characteristic heights of filament eruptions reported in observational studies.

In summary, our main found are following: Relative magnetic helicity, magnetic energy are predominantly concentrated at lower coronal heights. We recommend an extrapolation height of at least 81 Mm, as this not only encompasses complex magnetic structures such as filaments, but also achieves approximately 97\% numerical accuracy in magnetic helicity magnitude, more than 97\% accuracy in magnetic energy calculations, while saving approximately 38\% of computational resources. For studies that do not require high numerical precision, the extrapolation height may be moderately reduced to further conserve computational cost. This paper addresses an important question in magnetic field extrapolation: to what height should the extrapolation be carried out to achieve sufficient accuracy in parameter calculation without wasting substantial computational resources for marginal gains of only 1\%–3\%. This finding will provide valuable guidance for future extrapolation studies, particularly those involving large-sample active region analyses. In the future, we need to consider different extrapolation methods and magnetic helicity approaches, as well as conduct further research using different datasets.


\begin{authorcontribution}
S.Y. and Q.W. provided the code and guidance; H.X. assisted in organizing and provided the raw data; H.L. was responsible for part of the coding, data preprocessing, computations, plotting, and drafting the initial manuscript; S.Y., H.X., Q.W., A.D., and P.L. contributed to reviewing and revising the paper.
 All authors have read and agreed to the published version of \mbox{the manuscript.}
\end{authorcontribution}

\begin{fundinginformation}
\begin{sloppypar}
This research is supported by the National Key R\&D Program of China No. 2022YFF0503800, 2021YFA1600500, 2022YFF0503001; the National Natural Science Foundation of China (grants No. 12250005, 12073040, 12273059, 11973056, 12003051, 11573037, 12073041, 11427901, 11572005, 11611530679, 12473052 and 62427804); the Strategic Priority Research Program of the China Academy of Sciences (grants No. XDB0560000, XDA15052200, XDB09040200, XDA15010700, XDB0560301, and XDA15320102); by the Chinese Meridian Project (CMP); and by China's Space Origins Exploration Program.
\end{sloppypar}
\end{fundinginformation}

\begin{dataavailability}
The original SMFT data and the data generated in this study are available from the authors upon request.
\end{dataavailability}

\begin{codeavailability}
The extrapolation code and magnetic helicity computation code used in this study are both publicly available from previously published work. Other processing codes are available from the authors upon request.
\end{codeavailability}

\begin{ethics}
\begin{conflict}
The authors declare that they have no conflicts of interest.
\end{conflict}
\end{ethics}


\bibliographystyle{spr-mp-sola}
\bibliography{bibliography}

\IfFileExists{\jobname.bbl}{} {\typeout{}
\typeout{****************************************************}
\typeout{****************************************************}
\typeout{** Please run "bibtex \jobname" to obtain} \typeout{**
the bibliography and then re-run LaTeX} \typeout{** twice to fix
the references !}
\typeout{****************************************************}
\typeout{****************************************************}
\typeout{}}

\end{document}